\documentclass[preprint,rsi,amsmath,amssymb]{revtex4-1}
\usepackage{graphicx}
\usepackage{dcolumn}
\usepackage{bm}
\usepackage[utf8]{inputenc}
\usepackage[T1]{fontenc}
\usepackage{mathptmx}
\usepackage{float}
\usepackage{xcolor}

\begin{document}
\title[Filtering Noise in Time and Frequency Domain]{Filtering Noise in Time and Frequency Domain for Ultrafast 
Pump-Probe Performed Using Low Repetition Rate Lasers}
\author{Durga Prasad Khatua$^{1,2}$}
\author{Sabina Gurung$^{1,2}$}
\author{Asha Singh$^{1}$}
\author{Salahuddin Khan$^{1}$}
\author{Tarun Kumar Sharma$^{1,2}$}
\author{J. Jayabalan$^{1,2}$}
\email{jjaya@rrcat.gov.in}
\affiliation{$^{1}$ Nano Science Laboratory, Materials Science Section, Raja Ramanna Centre for Advanced 
	Technology, Indore, India - 452013. \\
$^{2}$ Homi Bhabha National Institute, Training School Complex, Anushakti Nagar, Mumbai, India - 400094.}

\date{\today}
\begin{abstract}
Optical pump-probe spectroscopy is a powerful tool to directly probe the carrier dynamics in materials down to 
sub-femtosecond resolution. To perform such measurement, while keeping the pump induced perturbation to 
the sample as small as possible, it is essential to have a detection scheme with high signal to noise ratio. 
Achieving such high signal to noise ratio is easy with phase sensitive detection based on lock-in-amplifier 
when a high repetition rate laser is used as the optical pulse source. However such a lock-in-amplifier based 
method does not work well when a low repetition rate laser is used for the measurement. In this article, a 
sensitive detection scheme which combines the advantages of boxcar which rejects noise in time domain and 
lock-in-amplifier which isolates signal in frequency domain for performing pump-probe measurements using 
low-repetition rate laser system is proposed and experimentally demonstrated. A theoretical model to explain 
the process of signal detection and a method to reduce the pulse to pulse energy fluctuation in probe pulses 
is presented. By performing pump-probe measurements at various detection conditions the optimum condition 
required for obtaining transient absorption signal with low noise is presented. The reported technique is 
not limited to pump-probe measurements and can be easily modified to suite for other sensitive measurements 
at low-repetition rates.
\end{abstract}

\maketitle

\section{\label{sec:level1}Introduction}
Ultrafast optical pump-probe spectroscopy is a powerful tool to directly study the carrier dynamics in materials. 
In the simplest form of this technique, a high energy pump pulse, which has temporal width shorter than the 
response time of the material, is used to change the carrier distribution in the sample. This change will modify the 
optical response of the material\cite{auston2013ultrashort,sutherland2003handbook,shah2013ultrafast,rossi2002theory}. 
The change in the optical property of the sample is then measured using another time delayed short and 
low energy probe pulse. The ultrafast dynamics in the sample can be probed by varying the delay between the pump 
and probe pulses obtained from an ultrafast laser. Since optical pulses of temporal width as short as few tens of 
atto-seconds are available, it is possible to study any process which are longer than the pulse width. 

To correctly understand the carrier dynamics, it is essential to keep the change induced by the pump to be 
as minimum as possible\cite{shah2013ultrafast}. Due to this the typical change in transmission or reflectivity
of the sample is expected to be very small ($\sim 10^{-3}$\% to $\sim 10^{-7}$\%). Hence such measurement
needs a high signal to noise (S/N) ratio\cite{PhysRevLett.65.764, rosker1986femtosecond}. Different types of 
noises can affect the measurements and reduce the S/N ratio. Typical sources of optical noise in such measurement 
are the fluctuations in the average power of the laser used in the measurement and any other background light 
which is being detected by a photodetector\cite{jonas1996pump, stevens2006signal}. In addition to these optical 
noises, the photodetector or the electronics used for the detection can pickup electrical noises from various 
sources\cite{flinn1968extent, milotti20021, regtien2018sensors}. Since the AC power supply has frequency 50 Hz, 
the electronic noise will be generated at this frequency. Further ordinary laboratory light sources would produce 
light at frequency 100 Hz (double of 50 Hz). The $1/f$ electrical noise will dominate at low frequencies. High 
frequency switches and RF sources will produce electrical pickup noise in high frequencies. Further, when a 
ultrafast laser pulse is incident on the photo-diode, there is a sudden increase in current flow which can be 
seen at the output of the photo-diode as an immense peak\cite{quinlan2013analysis}. Although such noises can 
cause random fluctuation in the signal at any given instant of time, a spectral analysis of the noise would 
show that each source will contribute to noise at different frequencies and its harmonics. 

One of the commonly used technique for achieving high signal to noise ratio in the pump-probe measurement 
is the phase sensitive detection technique which uses a lock-in-amplifier \cite{neelakantan1980signal, 
meade1983lock}. In this technique, first a frequency at which the noise is least is chosen, which typically
lies in few kHz range. The pump beam should be modulated at that frequency. The lock-in-amplifier then measures 
the signal at the modulation frequency thus isolating other noises which are at different frequencies resulting 
in a good signal to noise ratio\cite{meade1983lock, van2014response}. Most of the femtosecond oscillators operate 
at high repetition rates (for example $\sim$ 80 MHz). In such cases mechanical choppers or electro-optic modulators
operated at few kHz or few tens of kHz could be used to provide a good signal to noise ratio in the measurement. 
Typical femtosecond oscillators provide very low output per pulse energies and the pulse to pulse separation is also 
very short (12.5 ns for a 80 MHz system). In many samples the carrier relaxation and heat dissipation takes much 
longer time than the pulse to pulse separation in oscillators. Femtosecond amplifier systems can provide much 
higher per pulse energies with long pulse to pulse separation (1 ms for 1 kHz operation). 
If such amplifier systems are used for pump probe studies the modulation 
frequency have to be kept below 1 kHz. For obtaining good S/N ratio, it is essential to have good separation 
between the laser repetition rate and the chopping frequency. Too much reduction of chopping frequency is 
undesirable because at very low chopping frequencies $1/f$ noise will start dominating\cite{flinn1968extent, 
	milotti20021}. On the other hand the high-voltage switching circuits operating at the repetition rate of 
the laser system also produces electrical pickup noise in the detection system. In addition, the pulse to 
pulse energy fluctuation in case of 1 kHz systems are much more when compared to high repetition rate lasers. 
Thus, in case of low repetition rate lasers, signals are detected in time domain using boxcar or data 
acquisition cards with lot of averaging to achieve high S/N ratio \cite{collier1996low, bartolini2007real, 
anderson2007noise, werley2011pulsed, sahoo2018ultrafast}. Statistically, as the number of averaging ($N$) is 
increased, the standard deviation in the data can be reduced as $\sqrt{N}$ (for $N$ $>>$ 1). However, 
prolonged acquisition time can cause larger deviation in measurement conditions like temperature, sample 
condition and laser power. Such variations can lead to deviation in the measurement of the true optical response 
of the sample. Further, in many samples the decay time also depends on the excited carrier density and hence 
the pump power\cite{tang2007tailoring, jayabalan2009transient, salahuddin-QBeating-APL-2014}. In such cases 
averaging over large number of data with larger deviation in pump power would result in a wrong estimation 
for the response time.

In this article, a new technique which combines the advantages of both boxcar and lock-in-amplifier has been 
proposed and experimentally demonstrated by performing transient absorption measurement on a standard sample 
using a 1 kHz laser in two-color pump-probe geometry. We also analyze the problem theoretically to explain the 
process of noise elimination. In this technique, first a boxcar is used to detect the signal reducing the 
noise in time-domain. The output of the boxcar is then detected by a lock-in-amplifier to isolate noise in 
frequency domain thus improving S/N. The reported technique is general and will be useful for performing 
sensitive measurements at low-repetition rates.

\section{Theoretical Background}\label{Sec:Theoretical Background}
\begin{figure}[h]
	\begin{center}
		\centerline{\includegraphics[width=0.5\columnwidth]{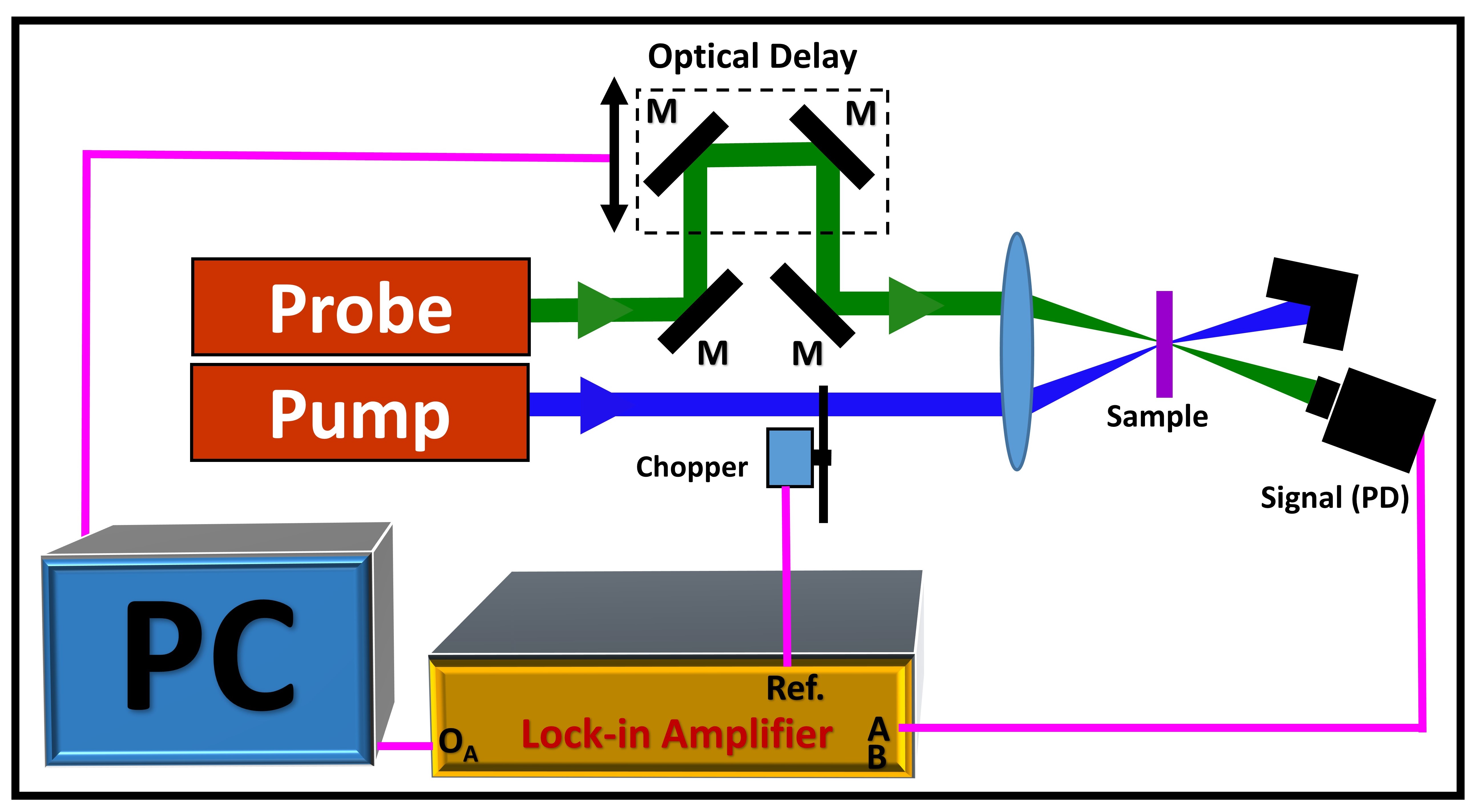}}
		\caption{ Schematic of a typical pump-probe setup used for the measurement of transient absorption based
			phase sensitive detection. M-Mirror and PD-Photodetector.}
		\label{Fig:TATheory}
	\end{center}
\end{figure}
Consider a standard transient pump-probe measurement performed on a sample using time synchronized pump 
and probe pulses (Fig.\ref{Fig:TATheory}). In such measurements the change in the absorption/reflectivity 
of the sample is first measured at a given delay between the pump and probe pulses and this measurement is 
then repeated at several other delays to obtain the complete temporal response\cite{norris2003femtosecond, stevens2006signal}. Let us consider a situation 
in which the laser source used in the measurement is highly stable and has sufficiently high pulse repetition 
rate. If the response time of the  photodetector (PD) being used is slow compared to the pulse to pulse 
separation, then the output voltage of the PD will be a constant which is proportional to the average power 
of the laser beam being detected. Let $P_{in}$ be the average power of the probe beam falling on the front 
side of the sample (for simplicity reflection losses are neglected in this analysis). The output voltage of 
the PD detecting the transmitted power through the sample can be written as,
\begin{align}
	V_O (t) = C T P_{in} + Re\left[ \sum_{\omega_N} V_N(\omega_N) e^{-i\omega_Nt} \right]
\end{align}
where $C$ is a constant which depends on the sensitivity of the PD and filters that are being used, $T$ is the 
transmission of the sample. $C$ will have the unit of A$^{-1}$. The second term accounts for the systematic and 
random electrical and optical noises in the measurement expressed in frequency domain. $V_N(\omega_N)$ is the amplitude of the 
noise at frequency $\omega_N$.  When the sample is excited by the pump pulse its optical response changes. For sufficiently 
small excitation conditions, the change in the absorption coefficient will be proportional to the third-order nonlinear 
absorption coefficient\cite{sutherland2003handbook, jayabalan2009transient}. Consider an instantaneously 
responding sample having a relaxation time 
much longer than the pulse width, the absorption coefficient after excitation by the pump pulse can be 
written as,
\begin{align}
\alpha(E_P, \tau) = \alpha_0 + \beta(\tau) E_P,
\end{align}
where $\alpha_0$ is the linear absorption coefficient, $E_P$ is the energy of the pump pulse, $\beta$ is related 
to the third-order nonlinear absorption coefficient with dimension of $1/(\alpha E_P)$ and $\tau$ is the time between the excitation of the pump 
pulse and arrival of the probe pulse. It is the temporal evolution of $\beta$ which finally reveals the carrier
dynamics in the sample. For a sufficiently small induced change in the optical response 
($\alpha_0 >> \beta E_P$), the transmission of the sample in presence of pump pulse is given by,
\begin{align}
T &= e^{-\alpha d}, \\
&\approx T_{0}\left[1-\beta(\tau) E_Pd\right].
\end{align}
where $T_0$ is the linear transmission of the sample having thickness $d$ and is equal to $\exp(-\alpha_0 d)$. The change in the 
transmission of the sample ($\Delta T = T- T_{0}$) induced by the pump pulse is given by,
\begin{align}
\Delta T (\tau) = -T_0 \beta(\tau) E_P d.
\label{Eq:DelTbyT}
\end{align}
If the pump beam is modulated at a frequency $\Omega_c$, the transmission of the sample will also change at that 
frequency between $T_{0}(1-\beta E_P d)$ (when pump pulse is falling on the sample) and $T_0$ (when pump is 
blocked). To get a good S/N ratio the $\Omega_c$ has to be chosen such that the electrical and optical noise 
in the lab is least at this frequency. Thus the output voltage of the PD when the pump beam is modulated is given by,
\begin{align}
V_O (t) = C T_{0}\left[1-\beta(\tau) E_P \mathcal{G}(t) d\right] P_{in} + Re\left[\sum_{\omega_N} 
V_N(\omega_N)e^{-i\omega_Nt} \right].
\label{Eq:VoPD}
\end{align}
Here, $\mathcal{G}(t)$ is a the pump beam modulating function. 
Assuming a square wave modulation of the pump pulse, which is typical for the mechanical chopping of the 
pump beam, the modulating function $\mathcal{G}(t)$ can be expressed in terms of infinite sum of sinusoidal 
waves given by\cite{arfken1999mathematical},
\begin{align}
\mathcal{G}(t) = \frac{4}{\pi} \sum_{k = 1}^{\infty} \frac{\sin \left[(2k-1) \Omega_c t\right]}{2k-1}.
\label{Eq:SqWave}
\end{align}
In a phase sensitive detection technique the output of the PD is connected to the input of a lock-in-amplifier. Using the chopper
frequency as the reference signal, the lock-in-amplifier will measure the amplitude of the voltage at $\Omega_c$.  Substituting 
Eq.\ref{Eq:SqWave} into Eq.\ref{Eq:VoPD} the output of the lock-in amplifier, $V_O$, at the frequency $\Omega_c$ 
can be obtained and is given by,
\begin{align}
V_L (\tau) = -\frac{4C}{\pi} T_{0}\beta(\tau) E_P P_{in} d.
\label{Eq:OnlyLIAOutput}
\end{align}
As mentioned earlier the $\Omega_c$ is chosen such that there is negligible noise at that frequency, hence $V_N(\Omega_c)  
\approx 0$. Using Eq.\ref{Eq:DelTbyT} and Eq.\ref{Eq:OnlyLIAOutput} the transient transmission at the delay $\tau$ can be written 
as,
\begin{align}
\frac{\Delta T (\tau)}{T_0}  = \frac{\pi}{4C} \frac{V_L (\tau)}{P_{in} T_{0}}. \label{Eq:DelTbyTBeforeADC}
\end{align}
A direct measurement of transmitted probe power though the sample using the same PD under similar condition by blocking the 
pump beam would produce a constant voltage, $V_0 = CT_0P_{in}$, at the output of the PD. Using this Eq.\ref{Eq:DelTbyTBeforeADC}
can now be written as,
\begin{align}
	\frac{\Delta T (\tau)}{T_0}  = \frac{\pi}{4} \frac{V_L (\tau)}{V_0}. \label{Eq:DelTbyTBeforeADCn}
\end{align}
Thus by using the output of the lock-in-amplifier ($V_L$) at various delays and $V_0$ the transient transmission
 through the sample could be measured with high S/N ratio. 

Repetition rate of several femtosecond oscillators are of the order of 80 MHz, which corresponds to a pulse to pulse separation of 
12.5 ns. Typical PDs which has microsecond response times will give a DC output when irradiated with output of such high repetition
rate lasers. By chopping the pump beam at few kHz, which is well away from the laser repetition rate and normal sources of noises,
a good signal to noise ratio for the transient measurement could be obtained\cite{jayabalan2009transient,khan2014coherent}. 
As mentioned in the introduction, many samples need higher pump pulse energies for generating sufficient change in probe as well as 
long pulse to pulse separation for providing sufficient time for the sample to relax. Femtosecond amplifier systems can provide much
higher energies and are generally operated at $\sim$ 1 kHz which corresponds to a pulse to pulse 
separation of 1 ms. When pump-probe measurements are carried out using such beams, the separation between the chopping frequency 
and laser repetition rate is very less. Further $1/f$ noise, other unwanted electrical and optical noises become difficult to 
isolate when the chopper frequency is less than 1 kHz. When low repetition rate lasers are used for a pump-probe measurement
the true signal arrives at the photodetector only for a very short duration, while rest of the time (much longer than the
signal duration) the photodetector generates unwanted signal from dark current, stray light and other electrical noises. Such 
unwanted accumulation of noise can be removed if the signal is measured only at the time of pulse arrival and rest of the data 
are rejected. Boxcar is an electrical instrument which can detect the signal from a photodetector for a specific duration
and produce a corresponding DC voltage at its output. The boxcar can be triggered by using a voltage pulse and it is possible 
to set a delay between the trigger and the measurement of signal. Boxcar also has provision to average the measured signal.
P. Bartolini {\it et al.} reported a real-time acquisition system in which a balanced detector (for minimizing common-mode noise), 
a boxcar averager and a data acquisition board were used for pump-probe measurement using low-repetition rate laser
\cite{bartolini2007real}. For pump-probe measurements with high-repetition rate laser, they have replaced the boxcar 
with a lock-in-amplifier. Acquisition of large amount of data using data acquisition cards and post processing of the 
collected information with lot of averaging has also been used for improving signal to noise ratio
\cite{bartolini2007real, anderson2007noise, werley2011pulsed, sahoo2018ultrafast}. 
Here, we present a new detection technique by utilizing the merits of both the boxcar and lock-in-amplifier instruments 
and the same is shown to be very useful in case of pump-probe experiments performed with a low repetition rate (1 kHz) 
laser system.

\section{Experimental Details}
\begin{figure}[h]
	\begin{center}
		\centerline{\includegraphics[width=0.5\columnwidth]{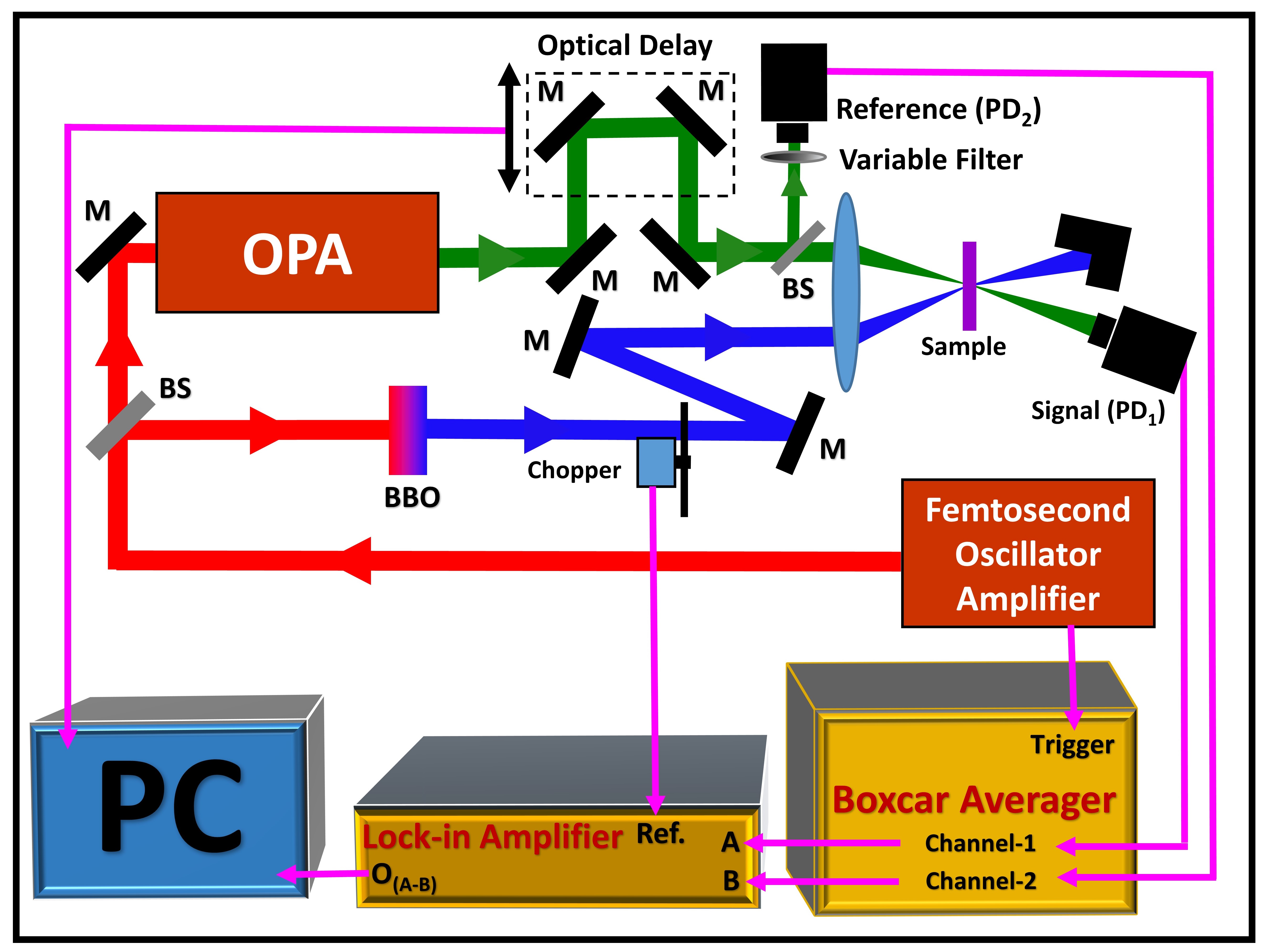}}
		\caption{ Schematic of the pump-probe setup along with the high sensitive detection system which combines advantages of
			boxcar and lock-in-amplifier. M-Mirror,  BS-Beam Splitter, BBO-Beta Barium Borate crystal and PD-Photodetector.}
		\label{Boxcarsystem}
	\end{center}
\end{figure}
For the demonstration of the present detection technique which combines the advantages of boxcar 
and lock-in-amplifier, pump-probe measurement 
was carried out on a Silver nanoparticle film by exciting at 400 nm and probing at 408 nm. Figure.\ref{Boxcarsystem} shows the 
schematic diagram of the experimental setup along with the details of the detection scheme. The schematic of the pump-probe setup 
remains same as that shown in Fig.\ref{Fig:TATheory} except for the detection of reference, but with a very different 
detection scheme which also incorporates a boxcar. 
The femtosecond oscillator-amplifier system delivers 35 femtosecond, 800 nm pulses at 1 kHz repetition 
rate. The 400 nm pump pulse was obtained by frequency doubling part of the 800 nm amplifier output using a Beta Barium 
Borate crystal (BBO). Another part of the 
amplifier output was used to pump an optical parametric amplifier (OPA). The typical standard deviation of the laser 
pulse-to-pulse energy fluctuation normalized to the average is nearly 3.4\%. The output of the OPA when operated at 408 nm was 
used as the probe beam. The pump and probe pulses were spatially overlapped on the sample using a lens. The temporal delay 
between the pump and probe pulses is controlled by passing the probe beam through a optical delay line. The transmitted probe 
power is measured using a normal photodetector (PD$_1$). The peak change in the voltage pulse generated by the photodetector 
is proportional to the pulse energy. Due to the design of the PD$_1$, this voltage pulse decays nearly exponentially with a time constant 
of $\sim$ 20 $\mu$s. Using the trigger from the laser timing circuit and by suitably choosing the delay and width of gate pulse, a 
boxcar (Stanford Research System SR200 Series) is made to detect the peak change in the photodetector voltage. All the 
measurements were carried out by operating boxcar without any amplification and under no averaging condition. The output of the 
boxcar generates a corresponding constant voltage which is now proportional to the laser pulse power. This output is connected 
to the input of a lock-in-amplifier (Signal Recovery Model 7265). Under no averaging condition in the boxcar output will 
be a constant voltage which changes every 1 ms due to the 1 kHz repetition rate of the laser. If the laser pulse to pulse energy 
fluctuation is negligible, the output of the boxcar will be a constant which is now nearly independent of repetition rate of the 
laser. Thus by inserting a boxcar in between the output of the PD$_1$ and lock-in-amplifier an experiment performed using a 
low-repetition rate laser has been converted to a situation which is similar to the constant output obtained from 
a slow photodetector when exposed to a high-repetition rate laser. In addition, the boxcar can also remove any noise 
that is time shifted with respect to the pulse arrival, the sharp electrical pickup noise which occur due to the pulse 
switch out process in the amplifier. Further, the response time of PD itself would not matter since the boxcar can be 
made to pickup the signal at a specific time on the output of PD signal.

Now the pump beam is mechanically chopped at a specific frequency ($\Omega_c$). If the laser pulse arrives at the chopper when 
the chopper blade is partially blocking the beam path the pulse will get cut spatially creating unwanted changes in the pump pulse
energy and random beam profile changes. To reduce such noise the chopper is placed at a location where the diameter of the beam 
is least. The output of the boxcar is connected to the channel A of the lock-in-amplifier. The pump-induced change in the transmission of the sample 
will modulate the boxcar output at the chopper frequency which can now be detected by the lock-in-amplifier as described in section 
\ref{Sec:Theoretical Background}.

\begin{figure}[]
	\begin{center}
		\centerline{\includegraphics[width=0.75\columnwidth]{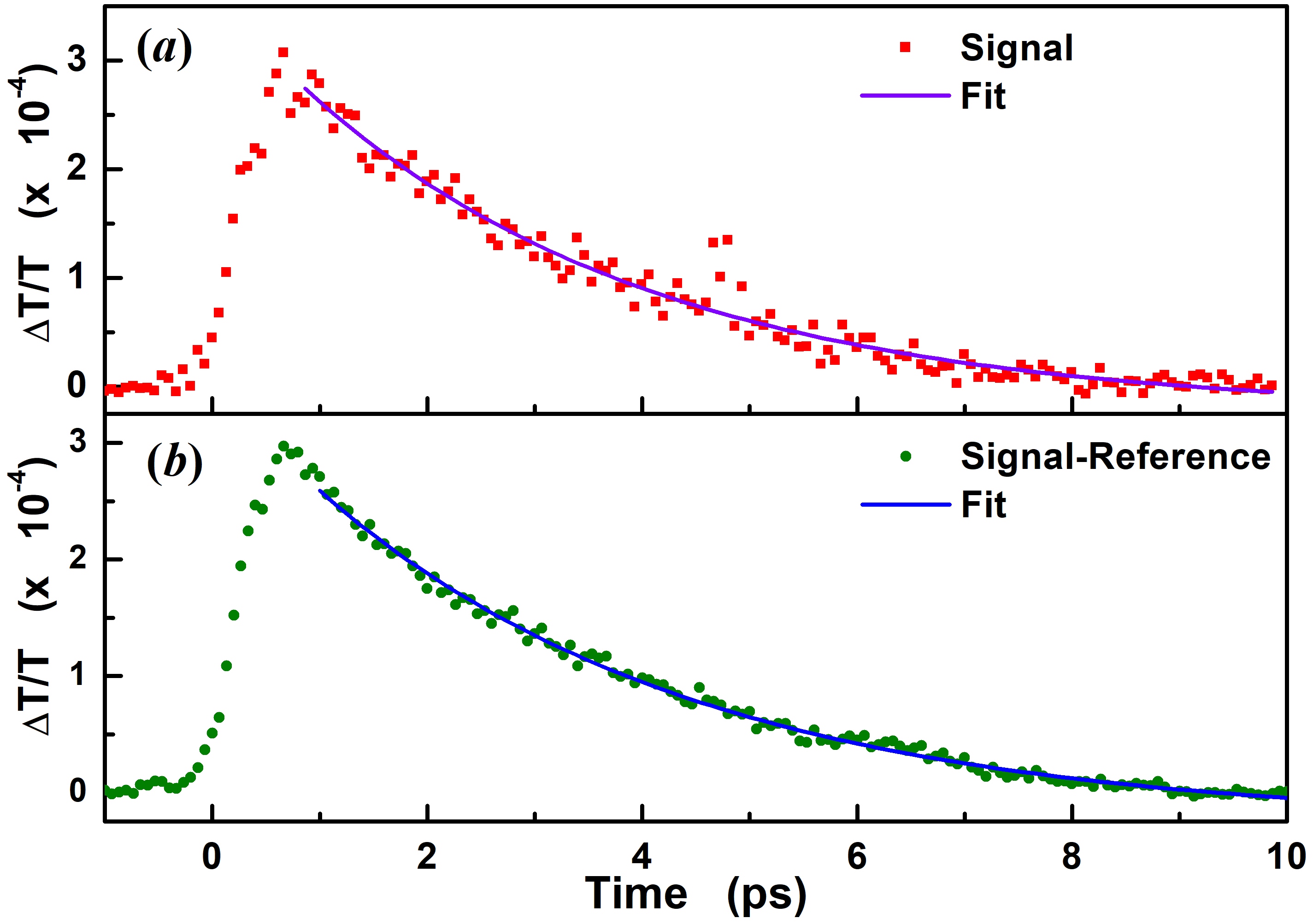}}
		\caption{ Transient transmission signal measured of a Ag thin particulate film. ({\it a}) the output of PD$_1$ is detected by boxcar 
			and the output of boxcar is then sensed with lock-in-amplifier at chopper frequency and ({\it b}) similar to ({\it a}) except that the fluctuation 
			in the signal is further reduced by subtracting a reference probe signal (see the text for details). }
		\label{Fig3-Transient}
	\end{center}
\end{figure}  

Figure \ref{Fig3-Transient} ({\it a}) shows the measured transient transmission of the Ag thin film sample performed  
by the present detector configuration. The chopping frequency was chosen to be 405 Hz. With the arrival of the pump 
pulse the transmission through the sample increases. In Ag nanoparticles the change in the transmission which 
is caused by the electron-phonon thermalization recovers in next few
picoseconds\cite{muskens2006femtosecond,jayabalan2009transient,singh2019counting}. A single exponential fit to the 
decay part of the signal gives a decay time of 3.3 $\pm$ 0.2 ps. Such transient response is typical for a metal 
nanoparticles. 

In order to study the statistical variation in the signal among various pump-probe detection parameters, the signal at the 
peak of the transient transmission ($\approx$ 0.5 ps) was recorded repeatedly. Then an estimate for the statistical 
variation ($S_D$) in percentage is calculated using,
\begin{align}
S_D = 100 \times \sqrt{\frac{1}{N}\sum_{i=1}^{N} \left(\frac{S_i}{S_{av}} - 1\right)^2 }, \label{Eq:SD} 
\end{align}
where $N$ is the number of data points, $S_i$ represents the individual measured data and $S_{av}$ is the average defined by,
 \begin{align}
S_{av} = \frac{1}{N}\sum_{i=1}^{N}S_i.
\end{align}
Since the data is normalized to the average, lower value of  $S_D$ implies a good S/N ratio. The $S_D$ estimated for the signal 
measured at the peak of $\Delta T/T$ shown in Fig.\ref{Fig3-Transient} ({\it a})) is 27 (N = 100).  

If the pulse to pulse power fluctuation in the laser is negligible, the collected data can be considered as similar to that 
of a standard transient transmission measurement performed using a high-rep rate laser. Generally the pulse 
to pulse energy stability of femtosecond amplifier system is much poorer when compared to that of oscillators.
Thus the transient signal measured using this low repetition rate oscillator-amplifier-OPA system 
will have additional noise due to pulse to pulse energy fluctuation. In the following we show, first theoretically and then 
experimentally the effect of such pulse to pulse energy variation in the probe pulse on the S/N ratio and a method 
to counter it. 

Let the average pulse energy of the probe pulse at the input of the sample be $E_{in}$ with a random pulse to pulse 
energy fluctuation $\Delta E_{in}$. The value of $\Delta E_{in}$ would change at every 1 ms in a 1 kHz repetition rate 
system. The distribution of $\Delta E_{in}$ for a ensemble of pulses would be a Gaussian, which is typical for the 
type of laser used in the present measurement. The voltage output of the boxcar channel which is detecting the transmitted 
probe pulse through the sample using PD$_1$ is, 
\begin{align}
V^{B}_O (t) = C^B T \left[ E_{in} + \Delta E^{in}\mathcal{H}(t)\right] + Re\left[ \sum_{\omega_N} V^B_N(\omega_N) 
e^{-i\omega_Nt} \right].
\end{align}
where $C^B$ is a constant which depends on the sensitivity of the PD$_1$ and gain in the boxcar. $V^B_N$ is the amplitude of 
electrical and optical noise at frequency $\omega_N$. Once again for small change in the transmission of the sample in 
presence of pump pulse ($E_P$) modulated by a function $\mathcal{G}(t)$, the output of the boxcar can be written as,
\begin{align}
V^B_O (t) = C^B T_{0} &\left[ 1 - \beta(\tau) E_P \mathcal{G}(t) d \right] \left[ E_{in} + \Delta E_{in}\mathcal{H}(t)
\right] \nonumber \\
&+ Re\left[ \sum_{\omega_N} V^B_N(\omega_N)e^{-i\omega_Nt} \right]. \label{Eq:BCOutput}
\end{align}
Here, $\mathcal{H}(t)$ is a unit step function which repeats with the arrival of probe pulse. Since 
$\Delta E_{in}$ varies randomly depending on the energy of each probe pulse, the product of  $\Delta E_{in}$ and
$\mathcal{H}(t)$ will be composed of several frequencies and can also have component at the chopper frequency. 

 For sufficiently small fluctuations in the probe energy $\Delta E_{in} << E_{in}$, the term containing the product 
 of $\beta$ and $\Delta E_{in}$ in Eq.\ref{Eq:BCOutput} can be neglected when compared to the other terms. Thus 
 the output of the lock-in-amplifier which is now detecting the BC output in channal A can be written as,
\begin{align}
V^B_L (\tau) = &- \frac{4C^B}{\pi}E_{in} T_{0}\beta(\tau) E_P d +  \Delta E_{in}\mathcal{F}(t) C^B T_{0} 
\label{Eq:LIAOutBCA}
\end{align}
where $\mathcal{F}(t)$ is the amplitude of Fourier transform of $\Delta E_{in} \mathcal{H}(t)$ at the chopping 
frequency and it depends on time because of the slow variations in laser pulse energy. Note that the second term 
in the Eq.\ref{Eq:LIAOutBCA} does not depend on the $\beta (\tau)$ of 
the sample, however depends directly on the pulse to pulse fluctuation. Thus, due to shot to shot energy fluctuation
in probe, additional noise (given by second term in Eq.\ref{Eq:LIAOutBCA}) will get added to the required signal, the 
first term. The $\Delta T/T$ presented in Fig.\ref{Fig3-Transient} will have fluctuations caused by the random probe 
energy variation. 

For removing this noise a small part of probe beam was reflected before the sample and was detected using another 
photodetector (PD$_2$). The output of the PD$_2$ is also processed by another channel of the boxcar. The output of the 
boxcar can be written as,
\begin{align}
V^{B2}_O (t) = C^{B2} T_f R \left[E_{in} + \Delta E_{in}\mathcal{H}(t)\right] + Re\left[\sum_{\omega_N} 
V_N^{B2}(\omega_N)e^{-i\omega_Nt} \right],
\end{align}
where $C^{B2}$  is a constant which depends on the sensitivity of the PD$_2$ and gain in boxcar. $R$ is the reflectance 
of the beam splitter used for obtaining the reference beam and $T_f$ is the transmission of filter used before 
PD$_2$. Using a variable filter, the $T_f$ was chosen such that $C^{B2} R T_f = C^B T_{0}$ (one can also change
the gain in the boxcar channel detecting the output of PD$_2$). This can be easily done directly by monitoring the boxcar 
outputs $V^B_O (t)$ and $V^{B2}_O (t)$ in an oscilloscope and making them almost equal. This voltage $V^{B2}_O (t)$ is 
then subtracted to the boxcar output of PD$_1$ ($V^B_O (t)$) electronically before feeding in to the lock-in-amplifier. 
Most of the commercial lock-in-amplifier do come with a provision for such subtraction (channel B) of voltage which could be 
directly used for this subtraction purpose. The output voltage after subtraction (A-B) can be written as,
\begin{align}
V^D_O (t) = - C^B T_{0} \beta(\tau) E_P \mathcal{G}(t) d  \left[ E_{in} + \Delta E_{in}\mathcal{H}(t)\right] \nonumber \\
+ Re\left[\sum_{\omega_N} (V^B_N - V^{B2}_N)e^{-i\omega_Nt} \right].
\end{align}
The output of the lock-in-amplifier which is now detecting the $V^D_O (t)$ using the chopping frequency as reference will be,
\begin{align}
V^B_L (\tau) = &- \frac{4C^B}{\pi}E_{in} T_{0}\beta(\tau) E_P d,
\label{Eq:VoutLIA}
\end{align}
A direct measurement of transmitted probe pulse energy though the sample using the same PD$_1$ under similar 
condition through the boxcar by blocking the pump beam would produce a constant voltage, $V^B_D = C^BT_0E_{in}$ 
at the output of the boxcar. Now using Eq.\ref{Eq:DelTbyT} and the measured $V^B_D$ the transient transmission through
the sample can estimated by,
\begin{align}
	\frac{\Delta T (\tau)}{T} = \frac{\pi}{4} \frac{V^B_L (\tau)}{V^B_D}.
\end{align}
It is important to note that in the present technique usage of an identical balanced detector is not essential as done 
by several earlier reports. Normal photodetectors can be used for measuring the reference and signal to reduce the noise
due to probe energy fluctuations. Only the DC output of boxcar needs to be made equal ($V^B_O (t)$ = $V^{B2}_O (t)$).

Figure \ref{Fig3-Transient} ({\it b}) shows the measured transient transmission signal of the metal thin film sample 
performed by the A-B configuration. A single exponential fit to the decay part of this signal gives a decay time of 
3.5 $\pm$ 0.1 ps. The $S_D$ of signal measured at the peak change in the present case is 10, which is nearly one 
third of that when the signal measured without subtraction of boxcar output of PD$_2$ from that of PD$_1$.

\begin{table}
	\caption{\label{Tab:TableCloseHz} The dependence of $S_D$ and peak to peak variation in the signal at different
		chopper frequencies $\Omega_c$ around 333 Hz.}
	\begin{ruledtabular}
		\begin{tabular}{ccc}
			$\Omega_c$ & $S_D$ (\%) & Peak to Peak fluctuation (\%) \\
			\hline
			230 & 15.5 & 95.3\\
			280 & 13.9 & 56.6\\
			333 & 2.9 & 14.5\\
			380 & 11.5 & 44.6\\
			420 & 11.6 & 61.2\\
		\end{tabular}
	\end{ruledtabular}
\end{table}

\begin{figure}[h]
	\begin{center}
		\centerline{\includegraphics[width=0.75\columnwidth]{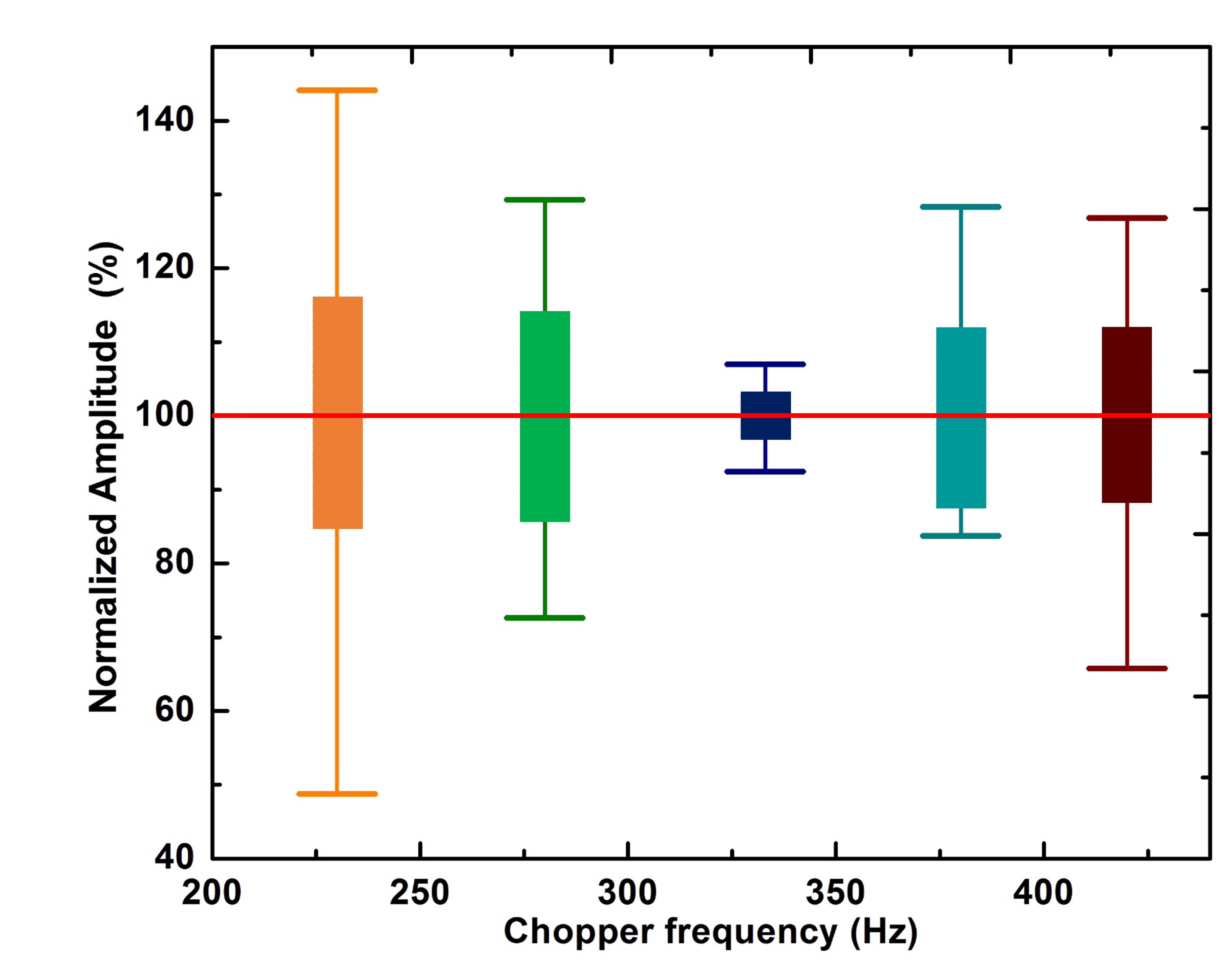}}
		\caption{The dependence of the fluctuations in the normalized signal measured at peak position  in A-B 
			configuration on the chopping frequencies around 333 Hz. The height of the rectangular box represents 
			the $S_D$ and the maximum and minimum signals values in the data set are also shown as bar.}
		\label{Fig:CloseFrequency}
	\end{center}
\end{figure} 

To further reduce the statistical fluctuations in the measurement, a chopper frequency dependent study was carried out 
on the same sample. The transient signal at the delay where peak occurs is recorded repeatedly in the A-B configuration 
($N$ is $\sim$ 100) at various chopper frequencies. The statistical variation in the peak signal as given by Eq.\ref{Eq:SD} 
was estimated at each chopper frequencies and are given in Table.\ref{Tab:TableCloseHz}. Further apart from $S_D$ it is 
also important to know about maximum and minimum signal obtained in the measured data to understand the stability in the
measured signal. Figure.\ref{Fig:CloseFrequency} shows the variation of $S_D$ on the chopping frequencies as a solid rectangle. The 
Fig.\ref{Fig:CloseFrequency} also shows the measured maximum and minimum signal values among the data set at each chopper 
frequency as bars. Clearly the $S_D$ is least for the chopping frequency 333 Hz. As the chopper frequency deviates away from 
333 Hz we find that the $S_D$ and the peak to peak variation in the signal also increases. 

\begin{table}
	\caption{\label{Tab:HarmonicChopper}The dependence of $S_D$ and peak to peak variation in the signal at different
		chopper frequencies which are factors of 1 kHz.}
	\begin{ruledtabular}
		\begin{tabular}{ccc}
			$\Omega_c$ & $S_D$ (\%) & Peak to Peak fluctuation (\%) \\
			\hline
			200 & 12.0 & 52.8\\
			250 & 9.8 & 46.2\\
			333 & 2.9 & 14.5\\
			500 & 12.0 & 58.0\\
		\end{tabular}
	\end{ruledtabular}
\end{table}

\begin{figure}[h]
	\begin{center}
		\centerline{\includegraphics[width=0.75\columnwidth]{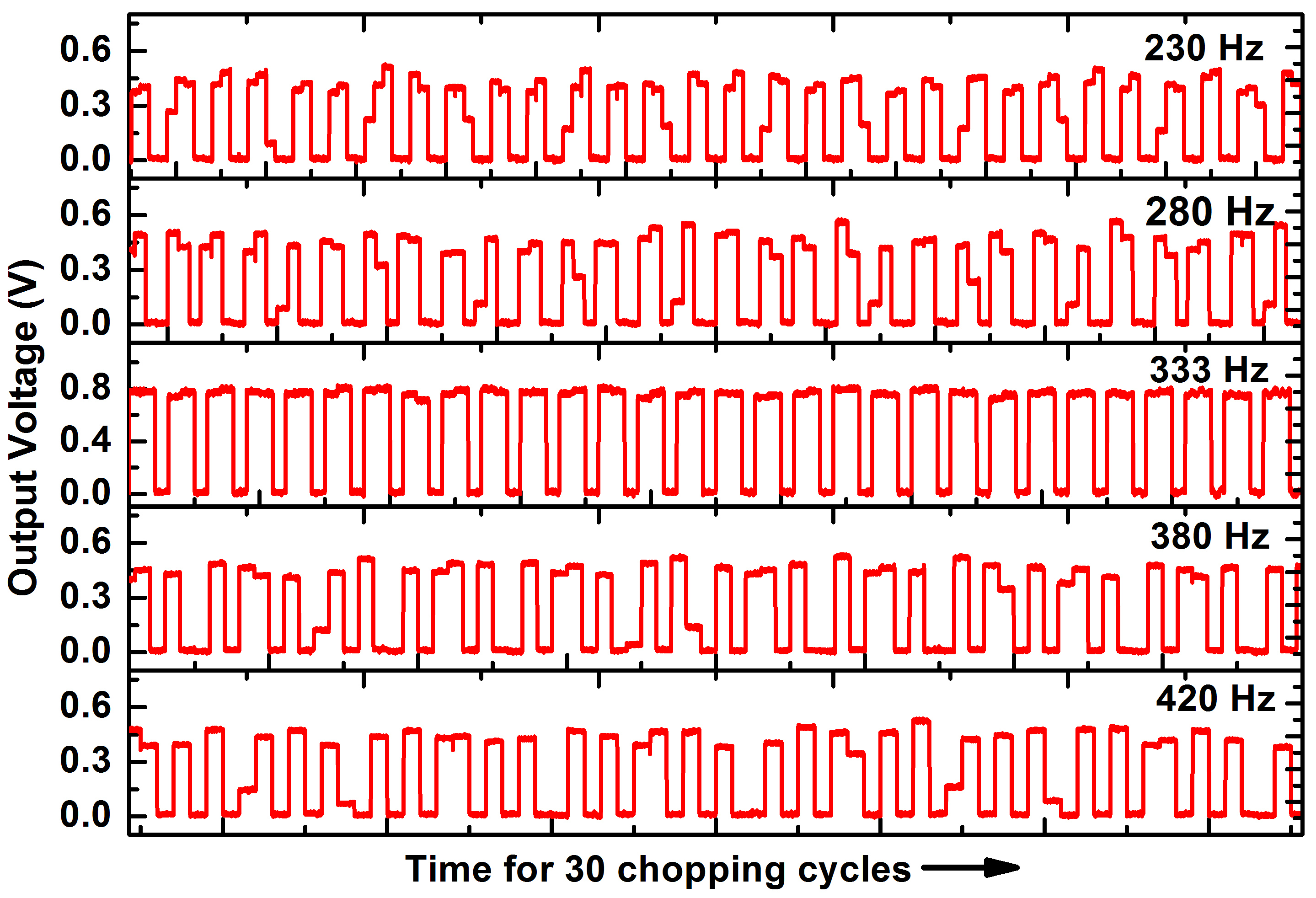}}
		\caption{Output voltage of the boxcar which is measuring the output of the photodetector sensing pump pulses passing though 
			the chopper operated at different frequencies. The time in the horizontal axis at each chopping frequency is chosen such that 
			there are 30 chopping cycles in view.}
		\label{Fig:CloseChopperHz-Time}
	\end{center}
\end{figure} 

When the laser pulse repetition rate is very large compared to the chopping frequency, the number of laser pulses blocked 
or unblocked in each cycle would remain nearly same. On the other hand, when the laser repetition rate and the chopper 
frequency is close an temporal aliasing process can lead to frequent changes in the number of pump pulses which are 
actually getting passed though the chopper. To measure the effect of chopper on the pump pulses we have directly 
measured the energy of the pump pulses after passing through the chopper. Fig.\ref{Fig:CloseChopperHz-Time} shows 
the output of the boxcar measuring the output of the photodetector detecting the modulated pump-pulses passing though 
the chopper for 30 cycles. Here zero amplitude means that the chopper had blocked the pulse while the higher amplitudes 
corresponds to the energy of the pulses passing through the chopper. Clearly the pump pulses are getting partially blocked by the 
chopper at frequencies other than 333 Hz. This phenomena can also be explained as follows. The chopper frequency 333 Hz 
is a factor of the laser repetition rate, 1 kHz. Thus in each chopper cycle two pulses are passed and one pulse is blocked and 
the process repeats (the reverse, two pulses are blocked and one pulse is passed is also possible depending on the starting 
time of chopper). On the other hand, if the chopper frequency is such that it is not a factor of laser repetition rate then 
frequently the chopper blade would be in a partially closed or partially open condition during the pulse arrival. Such partial 
cutting of pump pulse would lead to strong changes in the pump energy as well as spatial shape distortion of the beam 
profile at the sample place. Since the transient signal is proportional to the pump pulse energy  (see Eq.\ref{Eq:VoutLIA}), 
these variations would cause signal to fluctuate. Thus, 333 Hz is one of the possible factor of 1 kHz which can have least 
fluctuation in the signal as well as least variation in peak to peak. Hence, to avoid such chopping generated modulations
in the pump pulse it is essential chose the chopping frequency as one of the factor of the laser repetition rate.

\begin{figure}[h]
	\begin{center}
		\centerline{\includegraphics[width=0.75\columnwidth]{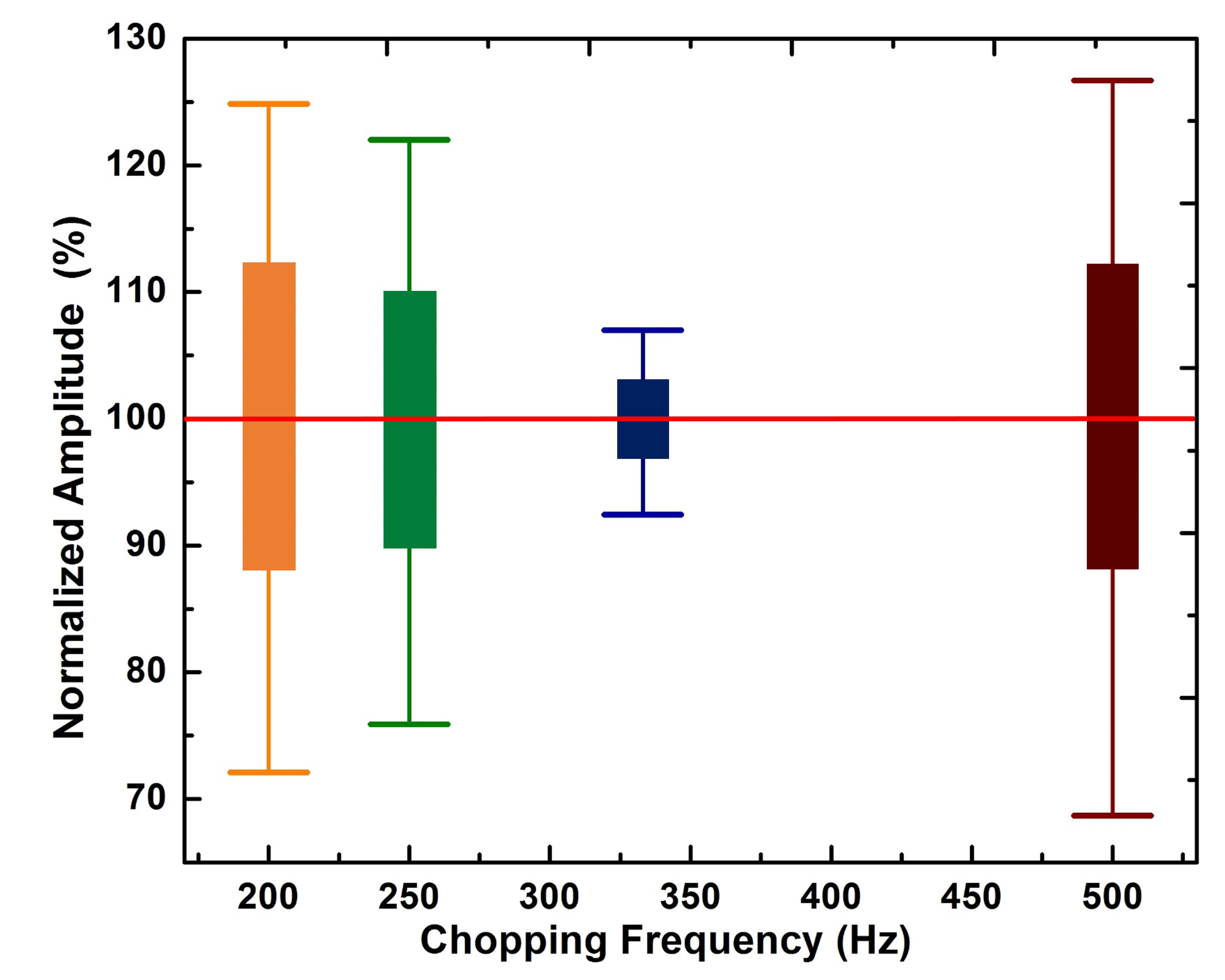}}
		\caption{The dependence of the fluctuations in the normalized signal measured at peak position  in A-B 
			configuration on the chopping frequencies which are factors of 1 kHz. The height of the rectangular box represents 
			the $S_D$ and the maximum and minimum signals values in the data set are also shown as bar.}
		\label{Fig:HarmonicFrequency}
	\end{center}
\end{figure}

\begin{figure}[h]
	\begin{center}
		\centerline{\includegraphics[width=0.75\columnwidth]{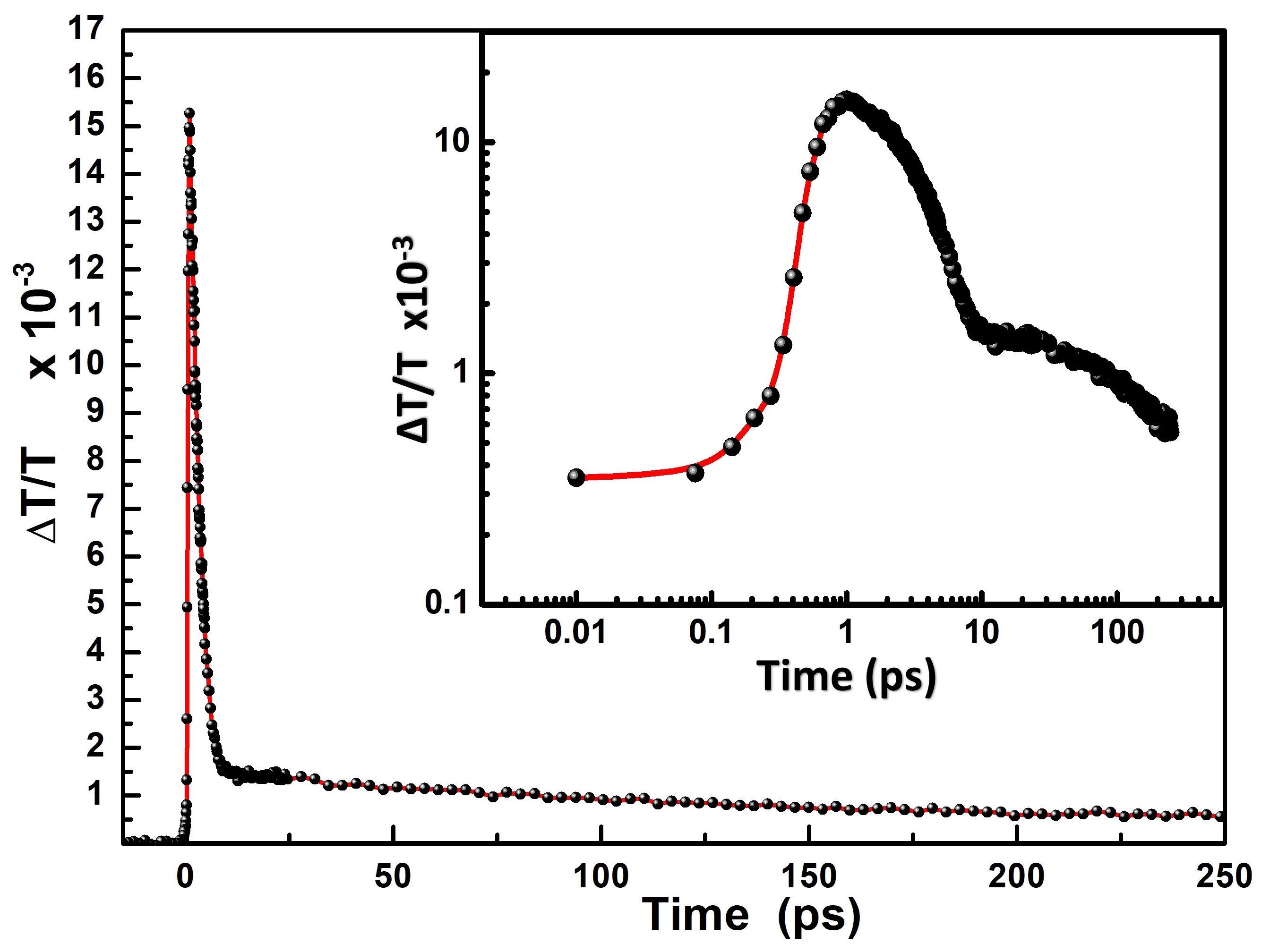}}
		\caption{A typical transient absorption signal measured with 333 Hz chopping. Inset shows the same curve in log-log scale.}
		\label{Fig:Transient333Hz}
	\end{center}
\end{figure} 

We have also performed similar measurement at all the chopper frequencies which are factors of 1 kHz in the A-B configuration. 
The statistical variation in the peak signal are given in Table.\ref{Tab:HarmonicChopper}. Figure.\ref{Fig:HarmonicFrequency} 
shows the dependence of $S_D$ and the measured maximum and minimum signal values on the harmonic chopping frequencies 
which are factors of 1 kHz. Once again the $S_D$ for the case of chopping frequency 333 Hz has the  least $S_D$. Although the 
chopper frequencies, 200 Hz, 250 Hz and 500 Hz are factors of 1 kHz still the noise at these frequencies are much higher than 
that of 333 Hz. Pump-probe measurements performed at very low chopper frequencies like 50 or 100 Hz (factors of 1 kHz) result 
in increasing the shot-to-shot fluctuations leading to even lower signal-to-noise ratio.
The AC power supply frequency which is being used in the experiment is 50 Hz and the on-off frequency of most 
of the laboratory light sources are driven by this AC power supply is 100 Hz. Thus, because of any kind of stray light present 
in the lab, it is well known that 50 Hz, 100 Hz and their higher harmonics are all not suitable for chopping frequencies. When 
chopped at these frequencies the lock-in-amplifier will sense these electronic 
noises or the other light which are being detected by the photodetectors. Since the chopping frequencies 200 Hz, 250 Hz and 
500 Hz are all higher harmonics of 50 Hz and 100 Hz the signal detected at these frequencies will have higher noise compare 
to that at 333 Hz. Several earlier reports used 500 Hz or various other chopping frequencies for the measurement of pump-probe 
signal with 1 kHz system\cite{gonccalves2016dual, werley2011pulsed, sahoo2018ultrafast}. 
Our result suggests that for a best S/N ratio, the pump should be chopped 
at 333 Hz. Further, even for the case where only lock-in-amplifier is used in the measurement the chopping has to be done at 
333 Hz. Although the technique reported here takes care of the fluctuations in data caused by the probe energy variation, 
still the variations in pump energy would cause the signal to vary (Eq.\ref{Eq:VoutLIA}). Figure.\ref{Fig:Transient333Hz} 
shows an transient absorption signal measured on a silver nanoparticle sample under fully optimized conditions and averaged 
over 8 measurements at each delay. In such long delay range a second slow decay in the transient signal which is due to the 
phonon-phonon interaction can also be clearly identified (see the inset of Fig.\ref{Fig:Transient333Hz})
\cite{muskens2006femtosecond,jayabalan2009transient}. 

\section{Conclusion}
In this work, a detection scheme which combines the advantages of boxcar and lock-in-amplifier for performing pump-probe using low-repetition rate laser system was demonstrated. A theoretical model was developed to explain the process of signal detection. The boxcar was operated such that it detects the peak signal from the photodetectors avoiding any noises that are being picked up by the 
photodetectors thus isolating noise in time domain. Chopping the pump and detecting the output of the boxcar using a lock-in-amplifier 
provides a way to isolate other electrical and optical noises that generally appear at other frequencies. In addition these noise 
isolations, a way to reduce the signal fluctuations due to pulse to pulse probe energy variation was also shown experimentally. 
A chopping frequency dependent study shows that for best signal to noise ratio the chopper frequency should be a factor of 
the laser repetition rate. Further, it should not also be a harmonic of the AC power supply. Thus in case of low-repetition rate
lasers, it is not only essential to move away from harmonics of AC supply, but also to a frequency which is a factor of its repetition rate.
Further, we also find that the gain in boxcar could also be used to increase the signal without much change in the S/N ratio.

\begin{acknowledgments}
Authors are thankful to Mr. Vijay Singh Dawar for his help during the experiments and Dr. Shweta Verma for providing the 
sample. Authors DP and SG are thankful to RRCAT, Indore under HBNI programme for providing the financial support.
\end{acknowledgments}

\nocite{*}
\bibliography{BoxcarLockin}

\end{document}